\documentclass[aps,prl,reprint,superscriptaddress,amsmath,amssymb]{revtex4-2}

\usepackage{graphicx}
\usepackage{listings}
\usepackage{longtable}
\usepackage{color}
\usepackage[colorlinks,allcolors=blue]{hyperref}
\usepackage[dvipsnames]{xcolor}

\definecolor{dkgreen}{rgb}{0,0.6,0}
\definecolor{gray}{rgb}{0.5,0.5,0.5}
\definecolor{mauve}{rgb}{0.58,0,0.82}
\newcommand\thefont{\expandafter\string\the\font}

\begin{document}

\title{Titanium Nitride Film on Sapphire Substrate with Low Dielectric Loss for Superconducting Qubits}

\author{Hao Deng}
\email{howl.dh@alibaba-inc.com}
\affiliation{Alibaba Quantum Laboratory, Alibaba Group, Hangzhou, Zhejiang 311121, China}
\author{Zhijun Song}
\affiliation{Alibaba Quantum Laboratory, Alibaba Group, Hangzhou, Zhejiang 311121, China}
\author{Ran Gao}
\affiliation{Alibaba Quantum Laboratory, Alibaba Group, Hangzhou, Zhejiang 311121, China}
\author{Tian Xia}
\affiliation{Alibaba Quantum Laboratory, Alibaba Group, Hangzhou, Zhejiang 311121, China}
\author{Feng Bao}
\affiliation{Alibaba Quantum Laboratory, Alibaba Group, Hangzhou, Zhejiang 311121, China}
\author{Xun Jiang}
\affiliation{Alibaba Quantum Laboratory, Alibaba Group, Hangzhou, Zhejiang 311121, China}
\author{Hsiang-Sheng Ku}
\affiliation{Alibaba Quantum Laboratory, Alibaba Group, Hangzhou, Zhejiang 311121, China}
\author{Zhisheng Li}
\affiliation{Alibaba Quantum Laboratory, Alibaba Group, Hangzhou, Zhejiang 311121, China}
\author{Xizheng Ma}
\affiliation{Alibaba Quantum Laboratory, Alibaba Group, Hangzhou, Zhejiang 311121, China}
\author{Jin Qin}
\affiliation{Alibaba Quantum Laboratory, Alibaba Group, Hangzhou, Zhejiang 311121, China}
\author{Hantao Sun}
\affiliation{Alibaba Quantum Laboratory, Alibaba Group, Hangzhou, Zhejiang 311121, China}
\author{Chengchun Tang}
\affiliation{Alibaba Quantum Laboratory, Alibaba Group, Hangzhou, Zhejiang 311121, China}
\author{Tenghui Wang}
\affiliation{Alibaba Quantum Laboratory, Alibaba Group, Hangzhou, Zhejiang 311121, China}
\author{Feng Wu}
\affiliation{Alibaba Quantum Laboratory, Alibaba Group, Hangzhou, Zhejiang 311121, China}
\author{Wenlong Yu}
\affiliation{Alibaba Quantum Laboratory, Alibaba Group, Hangzhou, Zhejiang 311121, China}
\author{Gengyan Zhang}
\affiliation{Alibaba Quantum Laboratory, Alibaba Group, Hangzhou, Zhejiang 311121, China}
\author{Xiaohang Zhang}
\affiliation{Alibaba Quantum Laboratory, Alibaba Group, Hangzhou, Zhejiang 311121, China}
\author{Jingwei Zhou}
\affiliation{Alibaba Quantum Laboratory, Alibaba Group, Hangzhou, Zhejiang 311121, China}
\author{Xing Zhu}
\affiliation{Alibaba Quantum Laboratory, Alibaba Group, Hangzhou, Zhejiang 311121, China}
\author{Yaoyun Shi}
\affiliation{Alibaba Quantum Laboratory, Alibaba Group USA, Bellevue, WA 98004, USA}
\author{Hui-Hai Zhao}
\affiliation{Alibaba Quantum Laboratory, Alibaba Group, Beijing 100102, China}
\author{Chunqing Deng}
\email{chunqing.cd@alibaba-inc.com}
\affiliation{Alibaba Quantum Laboratory, Alibaba Group, Hangzhou, Zhejiang 311121, China}

\begin{abstract}
Dielectric loss is one of the major decoherence sources of superconducting qubits.
Contemporary high-coherence superconducting qubits are formed by material systems
mostly consisting of superconducting films on substrate with low dielectric loss,
where the loss mainly originates from the surfaces and interfaces.
Among the multiple candidates for material systems,
a combination of titanium nitride (TiN) film and sapphire substrate
has good potential because of its chemical stability against oxidization,
and high quality at interfaces.
In this work, we report a TiN film deposited onto sapphire substrate
achieving low dielectric loss at the material interface.
Through the systematic characterizations of a series of transmon qubits
fabricated with identical batches of TiN base layers,
but different geometries of qubit shunting capacitors
with various participation ratios of the material interface,
we quantitatively extract the loss tangent value at
the substrate-metal interface smaller than $8.9 \times 10^{-4}$
in 1-nm disordered layer.
By optimizing the interface participation ratio of the transmon qubit,
we reproducibly achieve qubit lifetimes of up to 300 $\mu$s
and quality factors approaching 8 million.
We demonstrate that TiN film on sapphire substrate is an ideal material system
for high-coherence superconducting qubits.
Our analyses further suggest that the interface dielectric loss
around the Josephson junction part of the circuit could be the dominant limitation
of lifetimes for state-of-the-art transmon qubits.

\end{abstract}

\maketitle

\section{Introduction}

In superconducting quantum computing,
sufficiently long qubit coherence times compared to the time scales of operations
are the foundation of realizing practical quantum computation.
In the past two decades, the coherence times of superconducting qubits
have been improved by more than five orders of magnitude
through innovations in qubit design as well as the mitigation
of decoherence sources~\cite{kjaergaard2020superconducting}.
Among the diverse decoherence channels~\cite{Siddiqi2021},
dielectric loss has been indicated to be ubiquitous,
and has been the center of a few coherence-time breakthroughs
\cite{Barends.PRL.111.080502, Place.nat.comm.12.1779}
regarding the transmon platform.

The lifetime of a qubit, also known as the energy relaxation time $T_1$,
is directly related to the circuit quality factor
$Q = \omega_{\mathrm{q}} T_{1}$, where $\omega_{q}$ is the qubit frequency.
Regarding a transmon qubit, which is weakly anharmonic \cite{koch2007charge},
the dielectric-loss part of $Q$ can be decomposed into contributions
from various materials or regions as
\begin{equation}
Q^{-1} = \sum_{i} P_{i} \, \tan\delta_{i}.
\label{eq:total_q}
\end{equation}
Here, $i$ is the index indicating different spatial regions;
$P_{i}$ is the participation ratio (PR) of the $i$-th region,
representing the proportion of the electric energy stored in that part
to the total electric energy of the qubit;
and $\tan\delta_{i}$ characterizes the intrinsic dielectric loss of the $i$-th region.

It is intuitive that the region with the largest PR, usually,
the bulk of the substrate due to its large volume and dielectric constant,
would be the most crucial part of the dielectric loss.
However, previous studies show that,
the small-volume regions at the
interfaces of the qubit structure
contribute more significantly to dissipation
because of their large $\tan\delta_{i}$ values
\cite{Wang.APL.107.162601, Quintana.APL.105.062601, Sandberg.APL.100.262605,
Calusine.APL.112.062601, Melville.APL.117.124004, Woods.PRApp.12.014012}.
In a superconducting qubit,
the typical interface regions are the substrate-metal (SM),
substrate-air/vacuum (SA), and metal-air/vacuum (MA) layers illustrated in \autoref{fig:layout}(d).

To lower the dielectric loss,
$P_{i}$ and $\tan\delta_{i}$ are two independent variables that can be optimized.
However, because of the strong correlation between PR and qubit geometry
which will be discussed later,
the optimization of $P_{i}$ incurs potential trade-offs with regard to
the device structure, footprint size, and measurement configurations.
Therefore, reducing the $\tan\delta_{i}$ values of the interfaces
with respect to the material and fabrication,
is a general and fundamental approach
for reducing dielectric loss despite specific qubit designs.
In addition to the aluminum (Al) film on silicon substrate which is widely used in the community,
people actively explored other material systems
and the corresponding fabrication processes for low dielectric loss.
For example, titanium nitride (TiN) film on silicon substrate
has been explored in various devices
\cite{Sandberg.APL.100.262605, Chang.APL.103.012602,
Calusine.APL.112.062601, Woods.PRApp.12.014012, Melville.APL.117.124004}.
Recently, a study of tantalum film on sapphire substrate
reported low dielectric loss in a two-dimensional (2D) transmon qubit
\cite{Place.nat.comm.12.1779, Wang2022}.
Long-lifetime 2D transmon qubits fabricated with niobium film on silicon substrate
have been applied as sensitive probes
to characterize other dissipation mechanisms \cite{Gordon.ax.2105.14003}.
In these previous studies, a qubit quality factor in the 5-10 million range was achieved
by utilizing  high-quality material systems and optimizing the corresponding fabrication processes.
Further improvements in the qubit lifetime would require a quantitative understanding
of the loss mechanism in the circuits and its correspondence to the underlying material properties.

In this work, we report our study on
TiN film deposited onto sapphire substrate
as a material system providing low dielectric loss at the SM interface.
We select TiN as the superconductor because of
its chemical stability against oxidization,
and ability to form clean interfaces on certain substrates
\cite{Saha.JAP.72.3072, Krockenberger.JAP.112.083920, Richardson.JAP.127.235302}.
With regard to the substrate, sapphire is selected
for its highly passive surface, and extremely low dielectric loss \cite{Creedon.APL.98.222903}.
To characterize  quantitatively $\tan\delta_{\mathrm{SM}}$,
we systematically measure the $Q$ of 2D and three-dimensional (3D) transmon qubits
fabricated with identical batches of TiN base layers
depending on the $P_{\mathrm{SM}}$.
The value of $P_{\mathrm{SM}}$ covers a range of more than two orders of magnitude
by varying the design of the qubit shunting capacitor.
The experimental data from all the qubits are consistent with a single model
comprising only the interface dielectric loss of the TiN shunting capacitors
and that of the Josephson junction part of the circuits.
We extract a dielectric loss of the TiN film on sapphire substrate of $\tan\delta_{\mathrm{SM}}$
less than $8.9 \times 10^{-4}$ in a 1-nm-thick disordered layer
and a dielectric loss of the interfaces around the Al junctions of
$\tan\delta_{\mathrm{J}}$ $3.5 \times 10^{-3}$.
By optimizing the $P_{\mathrm{SM}}$ values of the transmon qubits,
we reproducibly achieve qubit lifetimes up to 300 $\mu$s
and quality factors approaching 8 million.

\section{Design and fabrication}

Regarding the interface region,
$P_{i}$ ($i = $ SM, SA, MA) can be described as~\cite{Wang.APL.107.162601, Calusine.APL.112.062601}
\begin{equation}
P_{i} =  \frac{1}{2} \epsilon_{i} \int_{0}^{t_i} dt \int_{S_{i}} |\bold{E}|^{2} \,ds / U_{\mathrm{tot}}
\label{eq:pi}
\end{equation}
where $t_{i}$, $\epsilon_{i}$, $S_{i}$, and $\bold{E}$ are
the thickness, dielectric constant, in-plane geometry,
and electric field of the $i$-th interface region,
and $U_{\mathrm{tot}}$ is the total electric energy of the qubit.
Usually, because the thicknesses of the interface layers
are much smaller than the dimensions of the qubit's in-plane geometry,
the electric field can be assumed to be uniform across these thicknesses.
Among the three regions listed above, the SM interface catches our attention:
on the one hand, the SM interface usually has major PR values
in the common designs of coplanar waveguide cavities and qubits
\cite{Wenner.APL.99.113513, Sandberg.APL.100.262605,
Quintana.APL.105.062601, Gambetta.IEEE.TransAppSc.27.1700205};
moreover, $\tan\delta_{\mathrm{SM}}$ is sensitive
to the selection of material and the method of fabrication,
and lacks well-tested post process
for further improvement, such as those applied on other interfaces
\cite{Quintana.APL.105.062601, Melville.APL.117.124004, Verjauw.PRApp.16.014018}.

\begin{figure}[tbp]
    \includegraphics[width = 1\columnwidth]{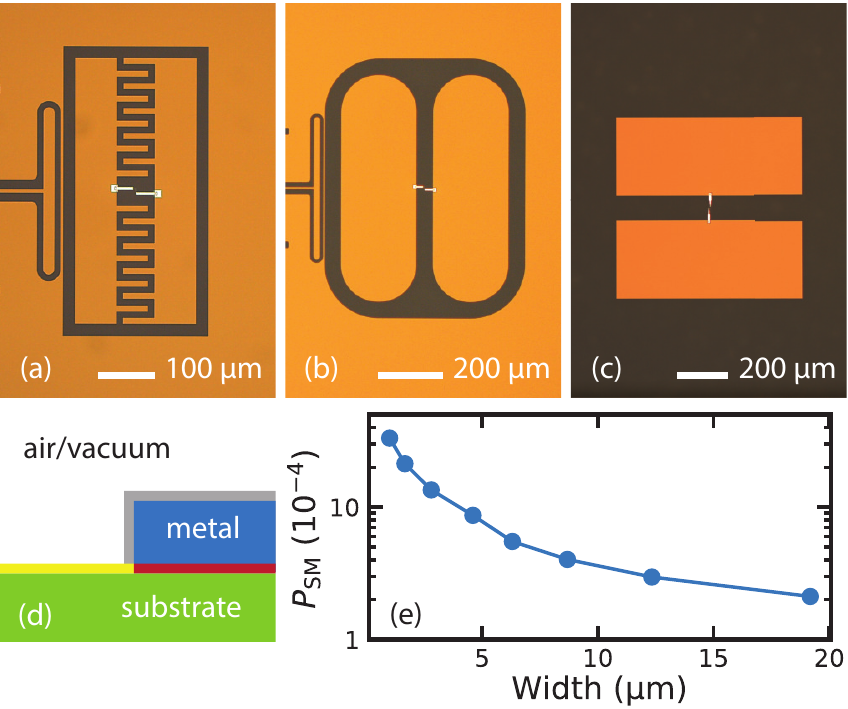}
    \caption{
    (a)-(c) Optical microscope images of typical interdigital 2D,
    dumbbell 2D, and dumbbell 3D transmon qubits in this study.
    The yellow area is the TiN base layer.
    The dark area is the sapphire substrate.
    The white area in the center of the image is the Al/AlO$_{x}$/Al Josephson junction.
    (d) Illustration of the cross section of the superconducting metal and substrate.
    The green, blue and white areas represent
    the substrate, superconducting metal film, and air/vacuum respectively.
    The yellow, red and gray regions indicate
    the potential lossy layers at the SA, SM, and MA interfaces.
    The illustration is not plotted to scale.
    (e) Dependence of $P_{\mathrm{SM}}$ on the width of gap and finger
    of the interdigital shunting capacitor,
    with $t_{\mathrm{SM}} = 1$ nm and $\epsilon_{\mathrm{SM}} = 10.15 \, \epsilon_{0}$.
    }
    \label{fig:layout}
\end{figure}

In \autoref{eq:pi}, $S_{\mathrm{SM}}$, the integral limit of the SM interface,
is determined by the geometry of the superconducting metal.
In our design of the transmon qubit,
we apply varied geometries of the qubit's shunting capacitor
in different regimes of $P_{\mathrm{SM}}$.
For the 2D transmon qubits with large $P_{\mathrm{SM}}$ values,
we select the interdigital geometry
which shows an ability to distribute more energy
in the SM interface region \cite{Gambetta.IEEE.TransAppSc.27.1700205}.
We keep the widths of the gap and finger identical, and vary them together
to reduce the number of variables in our design of the interdigital capacitor.
The amount and length of fingers are tuned correspondingly
to achieve the target values of capacitance.
Meanwhile, in the small-$P_{\mathrm{SM}}$ regime,
we use the dumbbell geometry of the shunting capacitor for the 2D transmon qubit.
The minimal $P_{\mathrm{SM}}$ in our study
is achieved by combining the dumbbell geometry with the 3D configuration.
The transmon qubits with varied geometries described above
are demonstrated in \autoref{fig:layout}(a)-(c).

For each geometry, we apply an independently developed electrostatic solver
to simulate the $P_{\mathrm{SM}}$ values of the varied shunting capacitor designs \cite{AQLTR.EMsolver}.
The other inputs required in the simulation
(i.e., $t_{\mathrm{SM}}$ and $\epsilon_{\mathrm{SM}}$) will be discussed later.
Focusing on the SM interface between the TiN film and sapphire substrate,
we do not include the contribution from the Josephson junction region.
The simulation result of the interdigital geometry plotted in \autoref{fig:layout}(e) shows that,
by varying the gap and finger widths in a range of 1-20~$\mu$m,
we can change the $P_{\mathrm{SM}}$ from approximately $2.1 \times 10^{-4}$ to $3.3 \times 10^{-3}$.
By extending the lower boundary of the $P_{\mathrm{SM}}$
with the dumbbell geometry and 3D configuration,
we achieve a series of transmon qubits covering a range of $P_{\mathrm{SM}}$ values of
more than two orders of magnitude,
which is beneficial for the fitting and analysis of the experimental data
to obtain a reasonable estimation of $\tan\delta_{\mathrm{SM}}$.

The transmon qubits in this study are fabricated in two steps,
namely the TiN base layer fabrication and the Josephson junction fabrication.
We first deposit the 100-nm-thick TiN films on the \textit{c}-plane sapphire substrates
via a magnetron sputtering system.
The deposited TiN films are then capped with SiN$_{x}$ layers
serving as inorganic hard masks \cite{AQLTR.TiNfilm}.
Patterns including the qubit shunting capacitors, the readout cavities, and the qubit drive lines
are first lithographically defined by a direct-write-laser system,
and then transferred to the SiN$_{x}$ hard mask layers via dry etching.
In the open regions of the hard masks, the TiN films are etched
in the SC-1 solution of RCA clean \cite{footnote}.
After the stripping of the SiN$_{x}$ hard mask layers with diluted hydrofluoric acid (HF),
the TiN base layers with the desired geometries are released.
In the next main step, we fabricate the Al/AlO$_{x}$/Al Josephson junctions
using the ``Manhattan'' technique \cite{Potts.IEEProce.SMT.148.225, Costache.JVSTB.30.04E105}.
The patterns of the junctions are defined by e-beam lithography,
followed by ion milling to remove the possible residuals from the development
and to clean the contact regions between the TiN base layers and Al leads.
By utilizing the shadowing effects of the e-beam resist patterns,
we sequentially deposit the Al leads of the junction in two orthogonal directions,
with an in situ oxidation process between the two evaporation steps
to form the AlO$_{x}$ insulating barriers.

\begin{figure}[tbp]
\includegraphics[width = 1\columnwidth]{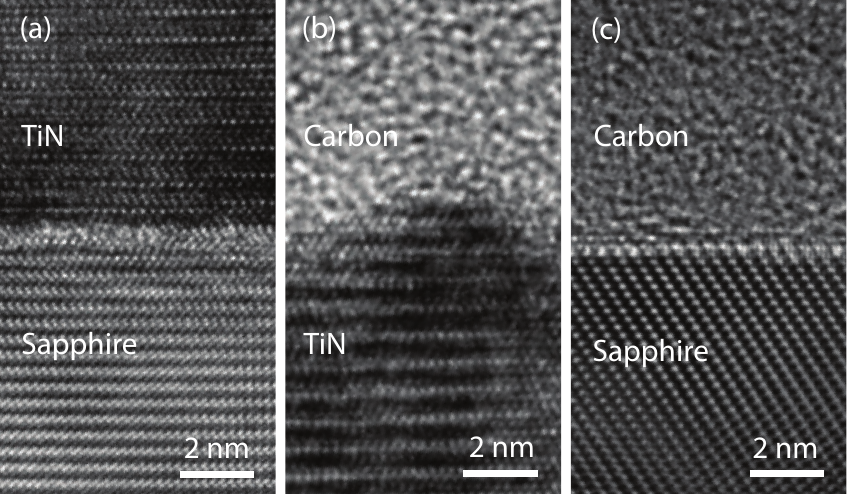}
\caption{
    (a)-(c) TEM images of the SM, MA, and SA interfaces
    of the TiN film on sapphire substrate.
    Note the disordered layer with a thickness of approximately 1 nm
    between the TiN film and sapphire substrate in (a).
    Although showing different contrast in (c),
    the atoms of sapphire at the SA interface
    generally locate at the expected lattice positions.
    In (b) and (c), the carbon capping layers are deposited
    for TEM sample preparation.
}
\label{fig:TiN_interface}
\end{figure}

To obtain the parameters required by the simulation of the $P_{\mathrm{SM}}$,
we characterize the cross section of the TiN film on the sapphire substrate
with transmission electron microscopy (TEM),
and demonstrate the results in \autoref{fig:TiN_interface}.
The image of the SM interface shows that,
there is a thin, disordered layer
between the highly crystalline TiN film and sapphire substrate.
By assuming that the dissipation at the SM interface is induced by this disordered layer,
its average thickness, approximately 1 nm, is adopted as $t_{\mathrm{SM}}$.
In the TEM image, the disordered layer shows a contrast similar to the sapphire substrate,
implying that the main component of this layer could be AlO$_{x}$.
Therefore, we use the value of sapphire's dielectric constant
($10.15 \, \epsilon_{0}$, where $\epsilon_{0}$ is the vacuum permittivity)
as the $\epsilon_{\mathrm{SM}}$.
As a typical value for many metal oxides,
an $\epsilon_{\mathrm{SM}}$ of approximately $10 \, \epsilon_{0}$
is widely adopted in the simulation of PR \cite{Sandberg.APL.100.262605}.

We also check the morphologies of the MA and SA interfaces with TEM,
as shown in \autoref{fig:TiN_interface}(b) and (c).
The images indicate that, the lattices of the crystalline TiN and sapphire
extend to their surfaces without an obvious disordered layer,
implying that the MA and SA interfaces
are expected to show insignificant contributions to the dielectric loss.
The high-quality MA and SA interfaces could benefit
from the process designs of TiN wet etching, SiN$_{x}$ layer stripping,
and the chemical stability of the TiN-sapphire material system.
Briefly, as the wet etchant of TiN is effectively a remover
for organics and certain metallic contaminants \cite{Kern.JES.137.1887},
etching TiN film with SC-1 solution could be accompanied by in situ decontamination.
In the subsequent SiN$_{x}$ layer stripping,
the HF-based process also provides an effective post cleaning
for the material stacks to further reduce the dielectric losses at the MA and SA interfaces
\cite{Melville.APL.117.124004, Verjauw.PRApp.16.014018}.
Moreover, the chemical stability of TiN guarantees
its compatibility with sufficient HF treatment,
and suppresses the regeneration of the lossy dielectric layer from oxidation.
Although disordered layers are not observed
at the MA and SA interfaces through TEM imaging,
their possible contributions to the dielectric loss are discussed later.

\section{Results and discussion}

\begin{figure}[tbp]
\includegraphics[width = 1\columnwidth]{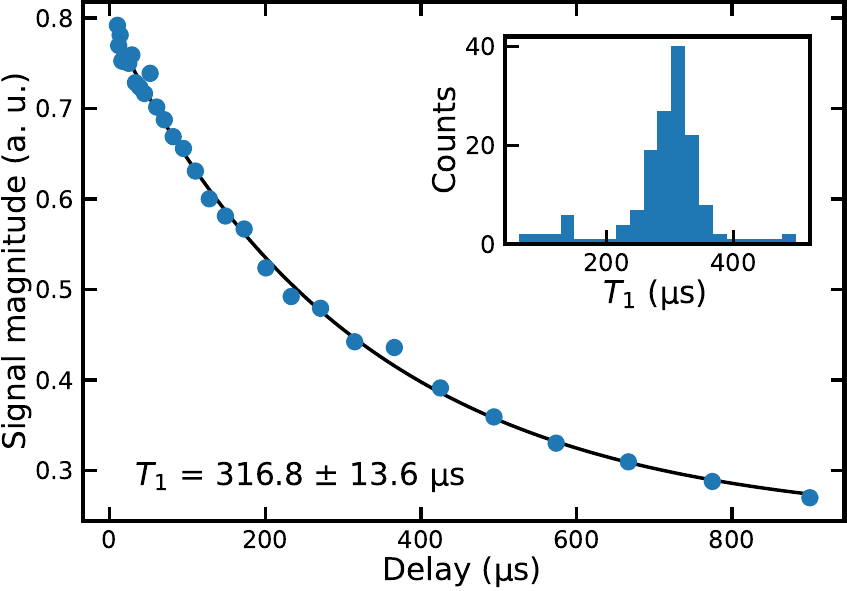}
\caption{
    Typical $T_{1}$ measurement result.
    The blue circles are experimental data points.
    The black curve is the fitting result of a single exponential decay.
    The fitting shows that $T_{1} = 316.8$ $\mu$s
    with a fitting error of $13.6$ $\mu$s.
    The inset shows the histogram distribution of $T_{1}$
    over approximately 100 rounds of continuous measurements.
}
\label{fig:t1}
\end{figure}

\begin{figure*}[tbp]
\includegraphics[width=2\columnwidth]{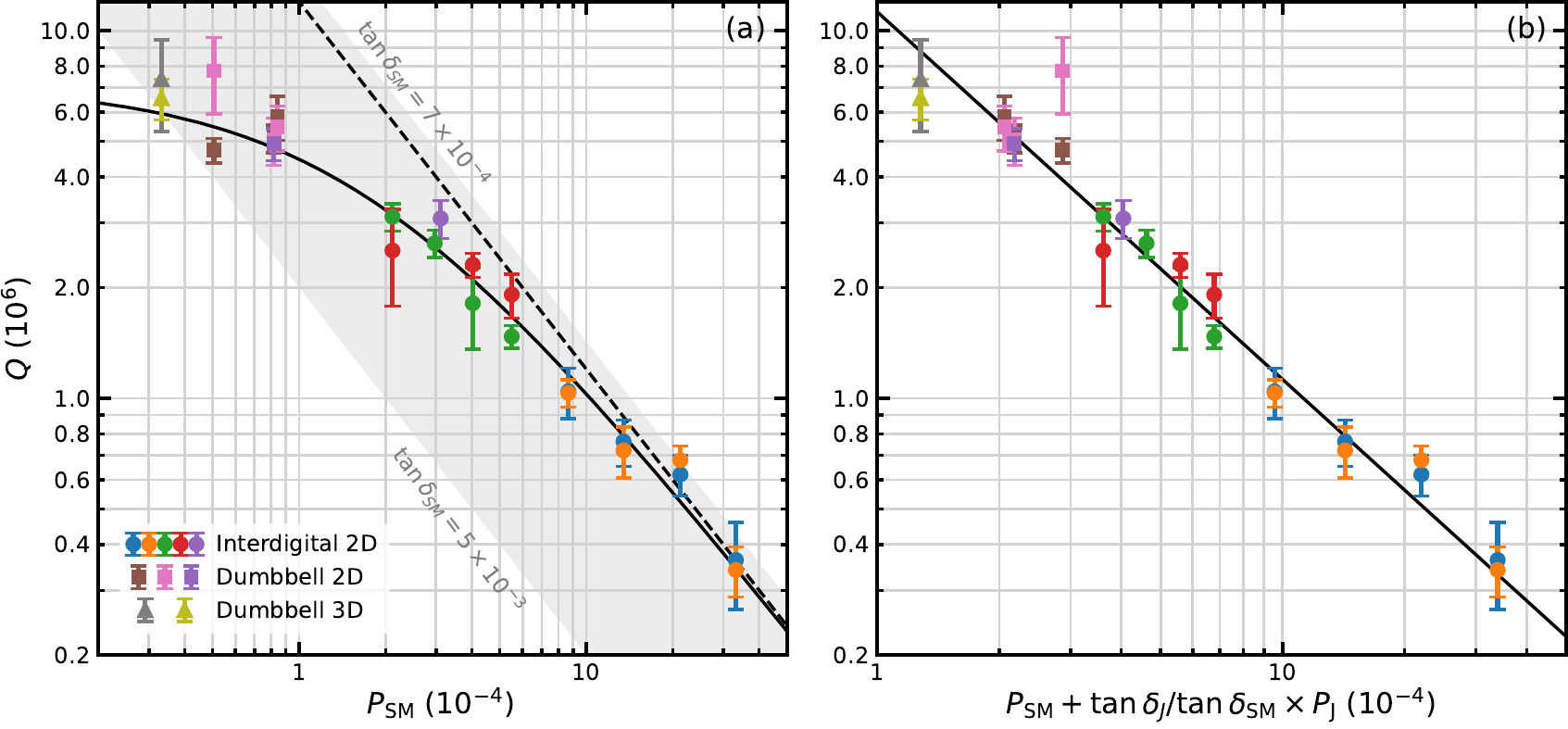}
\caption{
    (a) Dependence of $Q$ on $P_{\mathrm{SM}}$.
    The gray shaded zone indicates the limit of $Q$
    solely determined by the dielectric loss at the SM interface
    with the range of $\tan\delta_{\mathrm{SM}}$ annotated nearby.
    The black curve is the result of fitting
    with a model including both SM interface dielectric loss
    and a phenomenological parameter $Q_{0}$ representing extra dissipation.
    The black dashed line is the limit of $Q$
    determined by the dielectric loss at the SM interface
    with a $\tan\delta_{\mathrm{SM}} \simeq 8.3 \times 10^{-4}$ only.
    (b) Dependence of $Q$ on the normalized PR.
    The black line is the result of fitting
    with the model considering the dielectric loss
    depending on both $P_{\mathrm{SM}}$ and $P_{\mathrm{J}}$.
    In (a) and (b), the circles, squares, and triangles are the average $Q$ values
    of the interdigital 2D, dumbbell 2D, and dumbbell 3D transmon qubits, respectively.
    The error bars indicate the standard deviations.
    The dissipation from the Purcell effect induced by the readout cavity
    is subtracted from each data point.
    Different colors indicate different dies.
    One transmon qubit fails on the die represented by the red color.
}
\label{fig:q_psm}
\end{figure*}

We characterize our transmon qubits with the standard $T_{1}$ measurement \cite{footnote_2}.
To reduce the potential influence
of the temporal fluctuation of qubit performance
\cite{Klimov2018, burnett2019decoherence, carroll2021dynamics},
we repeat the measurement approximately 100 times,
and use the average $T_{1}$ as the representative value for data analysis \cite{footnote_3}.
The typical measurement result is demonstrated in \autoref{fig:t1}.
The relaxation of qubit generally shows an expected, single-exponential decay behavior,
and the statistic of $T_{1}$ has a relatively concentrated distribution.
In our systematic measurements,
13 designs of transmon qubits with different $P_{\mathrm{SM}}$ values on 5 kinds of dies are tested.
For each kind of die, multiple samples picked out randomly from the identical wafer
are measured to check the reproducibility \cite{footnote_4}.
On our best device, a time-averaged relaxation time $T_{1}$ of $291.7\pm 68.6~\mu$s is measured.
More details of the sample and measurement are available in the Appendix.

\autoref{fig:q_psm}(a) summarizes the experimental data of our measurements.
As a well-known extrinsic limit on $Q$ of qubits,
the dissipation from the Purcell effect induced by the readout cavity is subtracted
from each data point presented here \cite{footnote_5}.
We find that the measured $Q$ values increase monotonically with a decrease in $P_{\mathrm{SM}}$,
which is qualitatively consistent with the expectation of the dielectric loss at the SM interface.
However, there is an obvious deviation
between the experimental data
and the prediction made by a simple model
counting SM interface dielectric loss \textit{only}.
Although the data points in the large-$P_{\mathrm{SM}}$ regime
generally follow the straight line predicted by the simple model,
the $Q$ values in the small-$P_{\mathrm{SM}}$ regime
are systematically smaller than the expected values,
resulting in a variation in $\tan\delta_{\mathrm{SM}}$
between $7 \times 10^{-4}$ and $5 \times 10^{-3}$ (gray shaded zone),
depending on the choice of data points.
This observation implies that,
in addition to the readout-cavity-induced Purcell effect,
which is removed from each data point,
there is other dissipation mechanism
independent of the SM interface dielectric loss in the shunting capacitor region.
Following \autoref{eq:total_q}, we add a phenomenological parameter $Q_{0}$
and fit the data with $Q^{-1} = P_{\mathrm{SM}} \, \tan\delta_{\mathrm{SM}} + Q_{0}^{-1}$.
The fitting result is plotted in \autoref{fig:q_psm}(a) as the black curve,
which matches the experimental data
better than that from the simple model (black dashed line as an example).
Based on the patched model,
we get a $\tan\delta_{\mathrm{SM}} \simeq 8.3 \times 10^{-4}$
and a $Q_{0} \simeq 7.1 \times 10^{6}$.

Regarding the physical origin of the $P_{\mathrm{SM}}$-independent $Q_{0}$,
we hypothesize that it is correlated with the dissipation induced
by the Josephson junction regions.
Since all the qubits share a similar geometry of the junctions
not included in the simulation of $P_{\mathrm{SM}}$,
it is reasonable to expect that the energy loss induced by the junction region
is universal for all the transmon qubits in our measurements.
To verify this hypothesis,
we further simulate the PR of the junction region
by assuming that the dissipation there is still dominated
by the dielectric losses at the interfaces.
Similar to the PR simulation of the shunting capacitor region,
the thickness of the potential lossy layer
at each interface (i.e., SM, SA, and MA) is characterized by TEM \cite{footnote_6}.
The characterization shows that,
in addition to the approximately 1-nm-thick disordered layer at the SM interface,
the amorphous layer at the MA interface has a thickness of approximately 5.5 nm,
indicating a nonnegligible $P_{\mathrm{MA}}$.
Therefore, we count all the PR values in the junction region together as $P_{\mathrm{J}}$,
and characterize the intrinsic dielectric losses of the interface regions
with an overall $\tan\delta_{\mathrm{J}}$.
Then, we fit the data with
$Q^{-1} = P_{\mathrm{SM}} \, \tan\delta_{\mathrm{SM}} + P_{\mathrm{J}} \, \tan\delta_{\mathrm{J}}$,
and plot the fitting result in \autoref{fig:q_psm}(b) as the black line.
To demonstrate the data and fitting result clearly,
we plot $Q$ against the normalized PR defined by
$P_{\mathrm{SM}} + \tan\delta_{\mathrm{J}}/\tan\delta_{\mathrm{SM}} \, P_{\mathrm{J}}$ \cite{footnote_7}.
The analysis with the updated model gives a $\tan\delta_{\mathrm{SM}} \simeq 8.9 \times 10^{-4}$
and a $\tan\delta_{\mathrm{J}} \simeq 3.5 \times 10^{-3}$.

After considering the contribution from the junction region,
the data in \autoref{fig:q_psm}(b) show more consistent behavior
with the prediction from the dielectric loss model.
This result suggests that:
the quality of fabrication is reproducible among multiple samples,
and the impact from the temporal fluctuation on $T_{1}$
is suppressed by the repetition of measurements;
moreover, our simulation of the PR values of transmon qubits with varied geometries
and the assumption of the dielectric losses in the interface regions, are self-consistent.
With these well-controlled experiments,
we are able to achieve more direct characterizations
on the individual intrinsic dielectric losses of the TiN film for the shunting capacitors
and the Al film for the Josephson junctions on the sapphire substrate.

Regarding the fitting result of $\tan\delta_{\mathrm{SM}} \simeq 8.9 \times 10^{-4}$,
we would like to point out that, this value sets an upper limit for
the dielectric loss at the SM interface.
Our simulation reveals that,
$P_\mathrm{SA}$ and $P_\mathrm{MA}$ are approximately proportional to $P_{\mathrm{SM}}$
especially in the case of interdigital shunting capacitors,
consistent with previous studies
\cite{Wang.APL.107.162601, Gambetta.IEEE.TransAppSc.27.1700205}.
Therefore, the potential dielectric losses of the SA and MA interfaces
would be attributed to $\tan\delta_{\mathrm{SM}}$ in the fitting,
leading to a result larger than its real value.
The extracted value of $\tan\delta_{\mathrm{SM}}$,
although overestimated, indicates that
the quality of the SM interface of the TiN film
is comparable to those of the best reported material systems with low dielectric losses
\cite{Sandberg.APL.100.262605, Calusine.APL.112.062601,
Woods.PRApp.12.014012, Melville.APL.117.124004, Place.nat.comm.12.1779}.

Although obtained from the data points
distributing in a relatively narrow range of $P_{\mathrm{J}}$,
the dielectric loss of the junction region,
a $\tan\delta_{\mathrm{J}} \simeq 3.5 \times 10^{-3}$,
is consistent with previous studies on Al devices
fabricated with a lift-off process \cite{Wang.APL.107.162601, dunsworth2017characterization}.
We note that for long-lifetime devices that have a small $P_{\mathrm{SM}}$,
approximately 80\% of the relaxation can be attributed to the disordered layers
at the interfaces of the Al electrodes forming the junction,
which leaves considerable space for further improvement of the $Q$ of transmon qubits.

In conclusion,
we implement a material system of TiN film deposited on sapphire substrate
with low dielectric loss at the SM interface.
Through the measurements of a series of transmon qubits
with varied shunting capacitor designs and measurement configurations,
we systematically characterize the $Q$ values of qubits versus $P_{\mathrm{SM}}$,
and estimate the loss tangent at the SM interface.
The experimental data show that,
$\tan\delta_{\mathrm{SM}}$ is less than $8.9 \times 10^{-4}$
in a 1-nm-thick disordered layer,
accompanied by a dissipation induced by the junction region
with a $\tan\delta_{\mathrm{J}} \simeq 3.5 \times 10^{-3}$.
On the basis of the low dielectric loss of the material,
we achieve transmon qubits with $T_{1}$ values of up to the 300 $\mu$s level
and $Q$ values of approximately 8 million with reproducibility.
Our work indicates that, the TiN film deposited
by magneto sputtering on sapphire substrate
is an ideal material system for superconducting quantum computing,
which provides a low-loss platform for
the implementation of high-coherence qubits
and in-depth studies on other dissipation mechanisms in superconducting circuits.
Our analyses further suggest that the interface dielectric loss
around the Josephson junction part of the circuit
could be the dominant limitation of lifetimes for state-of-the-art transmon qubits.

\section{Acknowledgment}
We thank all the members of the Alibaba Quantum Laboratory
for their support of this work.
We thank Y. Lin (University of Science and Technology of China)
for the TEM characterization.

\section{Appendix}

\subsection{Fabrication}

The 100-nm-thick TiN films were deposited by using a high-throughput sputter system
equipped with a 4-inch radio-frequency (RF) sputter gun (CS-200, ULVAC Inc.)
using a 99.999\% purity titanium target.
The deposition power was set at 600~W and the Ar-N$_{2}$ (ratio 4:1) gas mixture used
for the reactive sputtering was held at a pressure of 0.5~Pa.
The substrates used for this study were 2-inch,
single-side polished, \textit{c}-plan sapphire (Suzhou RDMICRO Co. Ltd.).

The 100-nm-thick SiN$_{x}$ hard mask layers were deposited
via a plasma-enhanced chemical vapor deposition (PECVD) system (Haasrode C200A, LEUVEN Instruments Inc.).
The deposition was performed at 200$^{\circ}$C in a gas mixture of N$_2$, SiH$_4$, and NH$_3$.
The deposition pressure was maintained at 500 mTorr,
and the RF power was set at 100 W.

The lithography was implemented by a DWL 2000
direct-write-laser lithography system (Heidelberg Instrument GmbH).
S1813 photoresists (Advanced Materials Inc.) were first spin-coated
and soft-baked at 115$^{\circ}$C for 60~s.
After exposure, the coated wafers were developed
in an MIF319 developer (Kayaku Advanced Materials Inc.) for 60~s,
and cleaned in the running DI water for 120~s.

The lithography pattern was subsequently transferred to the SiN$_{x}$ hard mask layers
using an inductively coupled plasma (ICP) system (PlasmaPro100, Oxford Instruments).
The dry-etch process was performed at 20$^{\circ}$C
in a gas mixture of SF$_6$ and CHF$_3$.
The ICP power was set at 800~W and the bias power was set at 20~W.

The wet etch of the TiN films was performed in SC-1 solution
[i.e., a mixture of deionized (DI) water, ammonium hydroxide (29\% NH$_3$ by weight),
and hydrogen peroxide (30\%) at a volumetric ratio of 6:1:1].
The process was performed at 60$^{\circ}$C with added stirring.
After the completion of the wet-etch process,
the wafers were cleaned in DI water heated to 60$^{\circ}$C,
followed by cleaning with running DI water for 2~min.
The SiN$_{x}$ hard mask layers were then stripped off
in diluted hydrofluoric solution (5\%).

The fabrication of the Al/AlO$_{x}$/Al Josephson junctions was performed
using a conventional ``Manhattan'' technique.
First, the lithography of the mask structures was implemented
by a high-resolution e-beam lithography system (JBX 8100FS, JOEL Inc.).
The e-beam resist stack was composed of 950-nm PMMA (MicroChem Inc.)
on 200-nm MMA (Advanced Materials Inc.).
The exposure current was 2 nA and the beam dosages were 1750 and 1550 uC/cm$^{2}$
at the junction-area and the connection pad regions, respectively.
The deposition of the junctions was performed using an e-beam evaporator (MEB-600, CSWN).
Before the deposition, a gentle ion mill was applied for 30~s
to remove the residual resists in the mask openings.
The first 40-nm-thick Al layer was deposited at 6 \AA/s.
After 10~min of cooling, O$_{2}$ was injected and maintained at 160~Pa for 6~min
to form the AlO$_{x}$ insulating barrier.
The second 80-nm-thick Al layer was then deposited at a rate of 6.8 \AA/s rate
after a rotation of the sample stage in an orthogonal direction.
After junction fabrication, lift-off was implemented by 2 h of soaking in acetone,
and another 2 rounds of acetone cleaning and one isopropanol cleaning in a sonication bath.

\subsection{Interfaces at the Al/AlO$_{x}$/Al Junction Region}

\begin{figure}[tbp]
\includegraphics[width=1\columnwidth]{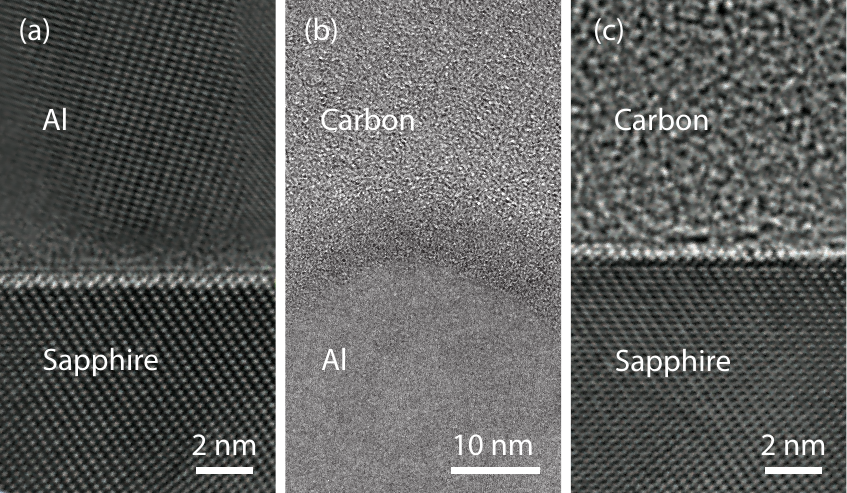}
\caption{
    (a)-(c) TEM images the of SM, MA, and SA interfaces
    at the Al/AlO$_{x}$/Al Josephson junction region.
    Note the disordered layers between
    the Al film and sapphire substrate in (a)
    and on the surface of the Al film in (b).
    The carbon capping layers are deposited to assist
    the TEM sample preparation in (b) and (c).
}
\label{fig:al_interface}
\end{figure}

Similar to the characterizations on the interfaces of
the TiN film on the sapphire substrate in the shunting capacitor region,
we perform TEM imaging on the cross section of
the Al film in the junction region,
and demonstrate the result in \autoref{fig:al_interface}.
The images of the interfaces show that, on average,
there are approximately 1-nm- and 5.5-nm-thick disordered layers
at the SM and MA interfaces, respectively,
which are adopted in the $P_{\mathrm{J}}$ simulation.
The dielectric constant applied in the simulation
is kept the same as the one assumed at the SM interface
of the shunting capacitor.

\subsection{Measurement}

\begin{figure}[tbp]
\includegraphics[width=1\columnwidth]{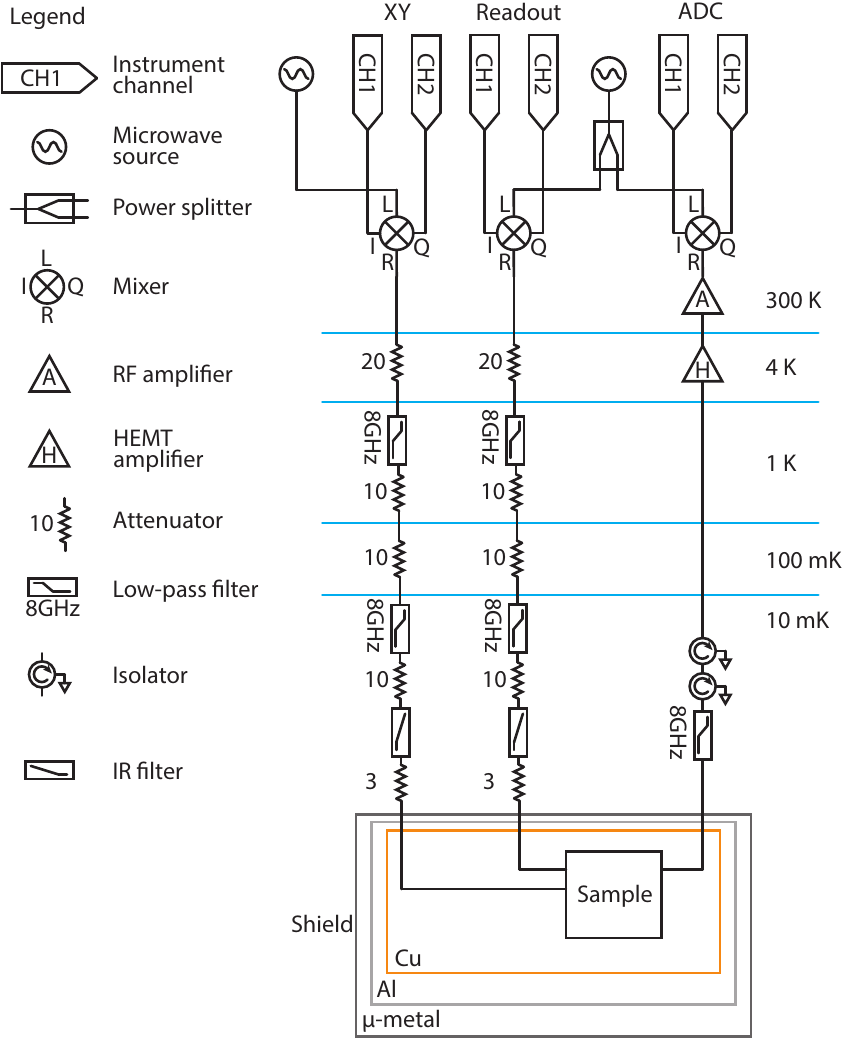}
\caption{
    Schematic of the measurement setup for each 2D transmon qubit.
    The attenuations (in units of dB) of the attenuators
    and the cutoff frequencies of the low-pass filters
    are annotated beside the legends in the schematic.
}
\label{fig:meas_setup}
\end{figure}

We use the standard heterodyne measurement setups
shown in \autoref{fig:meas_setup}
to characterize the $T_{1}$ values of our qubits.

We apply a HDAWG (Zurich Instruments Ltd.) and a UHFQA (Zurich Instruments Ltd.)
to generate qubit driving (XY) and readout signals, respectively,
at approximately $100$ MHz intermediate frequencies.
Then, these signals are modulated with carrier waves generated by
a microwave source, a SLFS 0211D signal generator (Sinolink Inc.),
through SIQM 0408 mixers (Sinolink Inc.).
On the signal input channels inside the dilution refrigerator,
cryogenic attenuators (XMA Inc.), F-30-8000-R low-pass filters (RLC Inc.)
and in-house developed infrared (IR) filters
are applied at different temperature stages,
to achieve thermal anchoring and noise suppression.

On the signal output channel,
F-30-8000-R low-pass filters (RLC Inc.) and LNF-ISISC4\_8A 2-stage isolators (Low Noise Factory Inc.)
are applied to reduce the possible noise and back reactions
from the upper stages at higher temperatures.
The output signal is amplified by LNC4\_8C high-mobility electron transistor
(HEMT) amplifiers (Low Noise Factory Inc.) at 4 K,
followed by additional amplification at 300 K with LNR4\_8C RF amplifiers (Low Noise Factory Inc.)
After the demodulation by a mixer with the identical carrier wave
divided from the one used in readout signal generation,
the output signal is sampled by
the analog-digital converter (ADC) of UHFQA (Zurich Instruments Ltd.).

To approach an ideal electromagnetic environment for our measurements,
we apply multiple-stage shields to isolate the samples from possible noise sources.
Copper (Cu), aluminum, and $\mu$-metal shields are applied from the inside out.
The channel feedthroughs on the shields are light-tight
to eliminate the influence of spray radiation from
high-temperature parts of the dilution refrigerator.
Moreover, we optimize the package of 2D transmon qubit samples
by hollowing the bottom stage under the die,
which has been indicated to be helpful in suppressing the coupling
between the package and the qubits \cite{Lienhard.IEEE.MTT.IMS.275, Huang.PRXQuantum.2.020306}.

Regarding the measurements of the dumbbell 3D transmon qubits,
we load our samples in rectangular Al cavities.
The setups are almost identical to those shown in \autoref{fig:meas_setup},
except that the XY and readout channels are merged by a power splitter at 300 K,
and share one signal input channel inside the dilution refrigerator.

\subsection{Qubit Parameters and Data Processing}

\begin{figure}[bp]
\includegraphics[width=1\columnwidth]{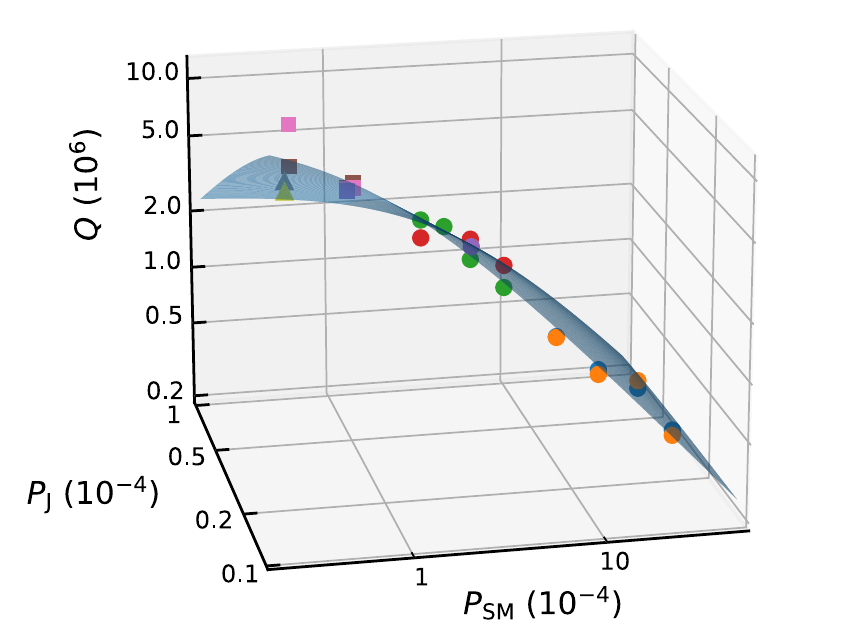}
\caption{
        Dependence of $Q$ on $P_{\mathrm{SM}}$ and $P_{\mathrm{J}}$.
        The semitransparent blue surface shows the fitting result
        of the model including the dielectric loss
        at both the SM interface of the TiN film and the Josephson junction region.
        Other legends are identical to those used in \autoref{fig:q_psm}.
    }
\label{fig:2d_fit}
\end{figure}

\begin{table*}[htbp]
\begin{ruledtabular}
\begin{tabular}{cccccccccc}
Device & Geometry & $\omega_{q}$ ($2\pi\cdot$GHz) & $\omega_{c}$ ($2\pi\cdot$GHz) & $g$ ($2\pi\cdot$MHz) & $T_{1}$ ($\mu$s) & $T_{\mathrm{Purcell}}$ (ms) & $Q$ ($10^{6}$) & $P_{\mathrm{SM}}$ ($10^{-4}$) & $P_{\mathrm{J}}$ ($10^{-4}$)\\
\colrule
D1-1 & Interdigital 2D & 4.43 & 6.46 & 37.3 & 36.8 (5.7) & 9.0  & 1.04 (0.16) & 8.67  & 0.22\\
D1-2 & Interdigital 2D & 4.04 & 6.39 & 36.4 & 29.8 (4.3) & 19.9 & 0.76 (0.11) & 13.49 & 0.19\\
D1-3 & Interdigital 2D & 4.47 & 6.31 & 36.8 & 22.2 (2.8) & 3.7  & 0.62 (0.08) & 21.27 & 0.18\\
D1-4 & Interdigital 2D & 4.35 & 6.26 & 37.1 & 13.3 (3.5) & 3.7  & 0.36 (0.10) & 33.19 & 0.19\\

D2-1 & Interdigital 2D & 3.83 & 6.48 & 35.6 & 42.7 (3.6) & 9.1  & 1.03 (0.09) & 8.67  & 0.22\\
D2-2 & Interdigital 2D & 3.94 & 6.41 & 36.0 & 29.0 (4.6) & 9.7  & 0.72 (0.11) & 13.49 & 0.19\\
D2-3 & Interdigital 2D & 3.90 & 6.33 & 37.6 & 27.7 (2.6) & 13.9 & 0.68 (0.06) & 21.27 & 0.18\\
D2-4 & Interdigital 2D & 3.81 & 6.28 & 37.6 & 14.2 (2.2) & 11.8 & 0.34 (0.05) & 33.19 & 0.19\\

D3-1 & Interdigital 2D & 4.21 & 7.18 & 38.4 & 116.3 (9.8) & 9.3  & 3.12 (0.27) & 2.12 & 0.38\\
D3-2 & Interdigital 2D & 3.80 & 7.15 & 40.1 & 109.1 (9.3) & 8.6  & 2.64 (0.23) & 2.97 & 0.41\\
D3-3 & Interdigital 2D & 4.02 & 7.10 & 39.8 & 71.3 (17.7) & 10.5 & 1.81 (0.45) & 4.03 & 0.39\\
D3-4 & Interdigital 2D & 4.34 & 7.35 & 41.9 & 53.6 (3.8)  & 7.9  & 1.47 (0.11) & 5.50 & 0.32\\

D4-1 & Interdigital 2D & 4.82 & 7.16 & 37.8 & 82.0 (23.8) & 5.7  & 2.52 (0.74) & 2.12 & 0.38\\
D4-3 & Interdigital 2D & 4.65 & 7.09 & 39.5 & 77.9 (5.8)  & 6.4  & 2.31 (0.17) & 4.03 & 0.39\\
D4-4 & Interdigital 2D & 3.70 & 7.34 & 42.7 & 81.5 (11.1) & 12.0 & 1.91 (0.26) & 5.50 & 0.32\\

D5-1 & Interdigital 2D & 4.11 & 6.74 & 39.7 & 191.5 (14.4) & 12.6 & 3.08 (0.37) & 3.11 & 0.23\\
D5-2 & Dumbbell 2D     & 4.32 & 6.69 & 44.0 & 181.5 (18.1) & 10.9 & 4.92 (0.49) & 0.82 & 0.34\\

D6-1 & Dumbbell 2D     & 4.06 & 7.42 & 47.2 & 185.6 (14.2) & 13.1 & 4.73 (0.36) & 0.51 & 0.59\\
D6-2 & Dumbbell 2D     & 3.78 & 6.96 & 46.4 & 245.4 (33.8) & 8.8  & 5.83 (0.80) & 0.84 & 0.31\\
D6-3 & Dumbbell 2D     & 4.45 & 6.72 & 46.5 & 182.5 (15.9) & 4.3  & 5.10 (0.44) & 0.82 & 0.34\\

D7-1 & Dumbbell 2D     & 4.24 & 7.38 & 49.5 & 291.7 (68.6) & 7.5  & 7.77 (1.83) & 0.51 & 0.59\\
D7-2 & Dumbbell 2D     & 4.10 & 6.92 & 46.9 & 212.4 (30.0) & 18.6 & 5.47 (0.77) & 0.84 & 0.31\\
D7-3 & Dumbbell 2D     & 4.31 & 6.68 & 46.4 & 186.1 (27.5) & 10.5 & 5.04 (0.74) & 0.82 & 0.34\\

D8-1 & Dumbbell 3D     & 5.32 & 7.50 & 85.4 & 197.2 & 0.8 & 8.82  & 0.33 & 0.24\\
D8-2 & Dumbbell 3D     & 5.38 & 7.50 & 85.6 & 186.8 & 0.5 & 10.31 & 0.33 & 0.24\\
D8-3 & Dumbbell 3D     & 5.65 & 7.48 & 84.5 & 148.5 & 0.6 & 7.13  & 0.33 & 0.24\\
D8-4 & Dumbbell 3D     & 5.23 & 7.49 & 84.6 & 179.4 & 1.9 & 6.52  & 0.33 & 0.24\\
D8-5 & Dumbbell 3D     & 5.22 & 7.51 & 84.1 & 196.5 & 1.8 & 7.26  & 0.33 & 0.24\\
D8-6 & Dumbbell 3D     & 5.43 & 7.48 & 84.1 & 114.9 & 1.5 & 4.25  & 0.33 & 0.24\\

D9-1 & Dumbbell 3D     & 4.55 & 7.49 & 85.7 & 203.8 & 3.2 & 6.23 & 0.33 & 0.24\\
D9-2 & Dumbbell 3D     & 4.70 & 7.51 & 86.3 & 236.0 & 2.7 & 7.63 & 0.33 & 0.24\\
D9-3 & Dumbbell 3D     & 4.70 & 7.48 & 85.0 & 218.5 & 6.0 & 6.70 & 0.33 & 0.24\\
D9-4 & Dumbbell 3D     & 4.85 & 7.48 & 85.3 & 179.0 & 4.9 & 5.66 & 0.33 & 0.24\\

\end{tabular}
\end{ruledtabular}
\caption{\label{tab:qp}
    Relevant parameters and experimental data of transmon qubits showed in \autoref{fig:q_psm}.
    $\omega_{q}$, $\omega_{c}$, and $g$ are the qubit frequency, readout cavity frequency,
    and the coupling strength between qubit and readout cavity.
    $T_{\mathrm{Purcell}}$ is the Purcell limit of qubit induced by the readout cavity.
    The standard deviations of $T_{1}$ and $Q$ are listed in parentheses.
    The experimental data of the dumbbell 3D transmon qubits (D8 and D9 series)
    are stable values from single-round measurements.
}
\end{table*}

In \autoref{tab:qp}, we list the relevant parameters
and experimental data of the transmon qubits reported in this work.

Following Ref. \cite{koch2007charge},
the Purcell limit induced by the readout cavity on the transmon qubit ($T_{\mathrm{Purcell}}$)
is estimated by $T_{\mathrm{Purcell}}^{-1} \simeq (g/\Delta)^2\kappa$,
where $g$ and $\Delta$ are the coupling strength and frequency difference
between the qubit and readout cavity, respectively,
and $\kappa$ is the line width of the readout-cavity spectrum.
$g$ is deduced from the dispersive shift of the readout cavity
$\chi \simeq g^2/\Delta$ in the dispersive regime.
We directly measure $\Delta$, $\kappa$, and $\chi$ in the experiments.
After estimating $T_{\mathrm{Purcell}}$,
we first subtract the contribution of the Purcell limit from each measured $T_{1}$,
then covert the result to $Q$, and finally apply statistics
to obtain the average value and standard deviation
shown in \autoref{fig:q_psm} and \autoref{tab:qp}.

The experimental data demonstrated in \autoref{fig:t1}
come from device D7-1 in \autoref{tab:qp}.
Regarding the two data points of the dumbbell 3D transmon qubits
in \autoref{fig:q_psm} and \autoref{fig:2d_fit},
the average $Q$ and standard deviation
come from the statistics of the single-round measurement results
of multiple devices fabricated in an identical batch,
listed as series D8 and D9 in \autoref{tab:qp}.

In \autoref{fig:2d_fit}, we show the original fitting result
of the model including the contributions
from both $\tan\delta_{\mathrm{SM}}$ and $\tan\delta_{\mathrm{J}}$.
The relative fitting errors are approximately 8\% and 5\%, respectively.
The expected $Q$ from the model with the fitted loss tangent values
is plotted as the semitransparent blue surface in the figure.

\bibliography{curr_lib}

\begin{thebibliography}{39}
\bibitem [{foo({\natexlab{d}})}]{footnote_4}%
  \BibitemOpen
  \href@noop {} {} ({\natexlab{d}}),\ \bibinfo {note} {one die is tested with
  single sample.}\BibitemShut {Stop}%
\end{thebibliography}

\end{document}